# A linear parameters study of ion cyclotron emission using drift ring beam distribution

Haozhe Kong[1], Huasheng Xie[2,3,*], Jizhong Sun[1,*]

[1]School of Physics, Dalian University of Technology, Dalian 116024, China
[2]Hebei Key Laboratory of Compact Fusion, Langfang 065001, China
[3]ENN Science and Technology Development Co., Ltd., Langfang 065001, China

E-mail: xiehuasheng@enn.cn and jsun@dlut.edu.cn

**Abstract**

Ion cyclotron emission (ICE) holds great potential as a diagnostic tool for fast ions in fusion devices. The theory of magnetoacoustic cyclotron instability (MCI), as an emission mechanism for ICE, states that MCI is driven by a velocity distribution of fast ions that approximates to a drift ring beam. In this study, the influence of key parameters (velocity spread of the fast ions, number density ratio, and instability propagation angle) on the linear MCI is systematically investigated using the linear kinetic dispersion relation solver BO (Xie H. 2019 *Comput. Phys. Comm.* **244** 343). The computational spectra region considered extends up to 40 times the ion cyclotron frequency. By examining the influence of these key parameters on MCI, several novel results have been obtained. In the case of MCI excited by super-Alfvénic fast ions (where the unique perpendicular speed of fast ion is greater than the perpendicular phase velocity of the fast Alfvén waves), the parallel velocity spread significantly affects the bandwidth of harmonics and the continuous spectrum, while the perpendicular velocity spread has a decisive effect on the MCI growth rate. As the velocity spread increases, the linear relationship between the MCI growth rate and the square root of the number density ratio transitions to a linear relationship between the MCI growth rate and the number density ratio. This finding provides a linear perspective explanation for the observed linear relation between fast ion number density and ICE intensity in JET. Furthermore, high harmonics are more sensitive to changes in propagation angle than low harmonics because a decrease in the propagation angle alters the dispersion relation of the fast Alfvén wave. In the case of MCI excited by sub-Alfvénic fast ions (where the unique perpendicular speed of fast ion is less than the perpendicular phase velocity of the fast Alfvén waves), a significant growth rate increase occurs at high harmonics due to the transition of sub-Alfvénic fast ions to super-Alfvénic fast ions. Similarly, for MCI excited by greatly sub-Alfvénic fast ions (where the unique perpendicular speed of fast ion is far less than the perpendicular phase velocity of the fast Alfvén waves), the growth rate at high harmonics also experiences a drastic increase compared to the low harmonic, thereby expanding the parameter range of the velocity spread.

## 1. Introduction

ICE, which was the first collective radiative instability driven by confined fusion-born ions in both JET [1,2] and TFTR [3] deuterium-tritium plasmas, has gained significant attention as a potential passive and non-invasive diagnostic technique for studying the population of fast ions in fusion devices [4-6]. Almost all tokamaks, including TFR [7,8], PDX [9], JT-60U [10-15], ASDEX-U [16-19], KSTAR [20,21], DIII-D [22-27], EAST [28-30], NSTX-U [31,32], TUMAN-3M [33,34], JET [1,2,35-38], TFTR [3,39], and HL-2A [40], as well as some stellarators such as LHD [41-45], W7-AS [46], and W7-X [47], have detected ICE. Experimental results have shown that the ICE intensity is nearly proportional to the neutron flux over a range of six orders of magnitude [1], and is closely related to various magnetohydrodynamic (MHD) activities, including edge localized mode (ELM) [1,4,20,21,29,30,48], fishbone mode [23], sawtooth ossifications [49], toroidal Alfvén eigenmode (TAE) [42], and plasma disruption [30]. These findings indicates that the excitation of ICE is inseparable from the presence of fast ions, which are generated by fusion reactions, ion cyclotron resonance heating (ICRH), and neutral beam injection (NBI) at the plasma center or boundary. Fast ions, which can be fusion products or minority species accelerated by ICRH, undergo radial drift excursions towards the outer edge plasma, resulting in a population inversion in velocity space that drives ICE. Similarly, fast ions injected by NBI also exhibit a population inversion near their injection point, contributing to the excitation of ICE. As a description of the population inversion of the above fast ions, drift ring beam distribution has been widely used in theories and simulations [1,2,14,38,44,50-68].

A successful theoretical explanation for the excitation mechanism of ICE is magnetoacoustic cyclotron instability (MCI) in the locally uniform approximation [53-57,69-73]. The MCI theory was first developed by Belikov and Kolesnichenko [71], in which the fast Alfvén and ion Bernstein branches at frequencies close to the cyclotron harmonics $l\Omega_F$ (where $l$ is the harmonic number) of a fast ion species F, are excited and propagate strictly perpendicular to the magnetic field in frequencies $\omega \gg \Omega_F$ (where $\Omega_F$ is the ion cyclotron frequency). Dendy et al. later extended this analysis to the frequency range $\omega \sim \Omega_F$, providing an explanation for ICE excitation [69]. The inclusion of finite parallel wave number $k_\parallel$ allowed MCI to being further successful in explaining the ICE excited by super-Alfvénic fast ions in JET [53]. Subsequently, MCI was used to account for ICE excited by sub-Alfvénic fast ions generated by fusion reactions and NBI [55]. Additionally, a variant of MCI successfully accounted for ICE excited by greatly sub-Alfvénic fast ions from NBI [54]. Analysis further considering magnetic field gradient and curvature drift effects predicted higher MCI growth rates [56,57].

However, the aforementioned MCI theory is valid when the instability growth rate $\gamma$ is larger than the inverse bounce/transit period of the fast ions $\tau_b^{-1}$ ($\gamma > \tau_b^{-1}$). When the instability growth rate $\gamma$ is lower than $\tau_b^{-1}$ ($\gamma < \tau_b^{-1}$), the analysis must consider toroidal effects and eigenmode structures [73-84]. In this work, our focus mainly lies on the locally uniform approximation MCI theory, where $\gamma > \tau_b^{-1}$. A large number of linear and nonlinear simulations have confirmed the validity of MCI in the locally uniform approximation [14,38,44,50-70,85], and these simulations capture most of the key observed features of the ICE measurements in JET and TFTR experiments, including the simultaneous excitation of all cyclotron harmonics, the splitting of spectral peaks, ICE intensity scaling approximately linearly with the fast ion number density, the strong growth rate for nearly perpendicular wave propagation, and the congruence between the linear theory and the observed signal intensities. The experimental phenomenon of the un-captured continuous spectrum for $l > 8$ in JET has also gained further understanding in this work. What needs to be emphasized is the congruence between linear theory and the observed signal intensities, including the linear simulation results agreeing surprisingly well with both the peaks in ICE intensity at ion cyclotron harmonics and the trend of increasing intensity with harmonic number [58,59], and a striking correlation between the time evolution of the maximum linear growth rate and the observed time evolution of the ICE amplitude [55,66]. Nonlinear simulations [58,59] suggest that MCI is intrinsically self-limiting on very fast timescales, providing an explanation for the observed correlation between linear theory and ICE intensity. Therefore, the self-limitation of MCI suggests that linear simulation is still a very important way to study ICE.

The key observed features of the ICE measurements captured by the simulations are closely related to the velocity spread of the fast ions, the number density ratio $\xi_F$ (the ratio of the fast ion number density to the background ion number density), and the instability propagation angle $\theta$ (the angle between the wave propagation and the ambient magnetic field), and these three parameters have been the focus of MCI research. However, previous linear and nonlinear simulations have not systematically studied the influence of these key parameters on MCI for the drift ring beam distribution in three cases: super-Alfvénic, sub-Alfvénic, and greatly sub-Alfvénic fast ions. This is a limitation for using ICE diagnosis to obtain fast ion information in future experiments. Therefore, this work provides a systematic investigation of the influence of these key parameters on linear MCI. The computational spectra region considered in this work is up to 40 times of the ion cyclotron frequency, which is rarely explored in other simulations. In addition, a more realistic experimental condition is considered in the simulation of ICE excited by super-Alfvénic fast ions. The present work is carried out by using the fully kinetic dispersion relation program BO (further details provided in Appendix A) [86-88]. The BO program has the advantage of providing all solutions of a linear kinetic plasma system without requiring an initial guess for root finding, which distinguishes it from previous simulation programs. This allows for parameter scanning across a wide range. Furthermore, the BO program model includes the wave electric field parallel to the ambient magnetic field. Therefore, MCI simulation carried out by using the BO program can be extended to arbitrary angle.

In section 2, the simulation results of BO program are successfully compared with the previous simulation results. In sections 3, 4 and 5, detailed simulation results are presented for MCI excited by super-Alfvénic, sub-Alfvénic, and greatly sub-Alfvénic fast ions, respectively. Conclusions are presented in section 6.

## 2. Benchmark

In this section, the simulation results of the BO program are successfully compared with the linear theory for super-Alfvénic and greatly sub-Alfvénic fast ions, respectively. In all our simulations, the velocity space distribution of the fast ions follows a drift ring beam distribution [86]:

$$f \propto \exp\left(-\frac{(v_\parallel - v_d)^2}{v_r^2}\right) \exp\left(-\frac{(v_\perp - u_\perp)^2}{u_r^2}\right).$$

Here $v_\parallel$ and $v_\perp$ denote velocity components parallel and perpendicular to the magnetic field, and $v_d$ and $u_\perp$ are constants that define the average parallel drift speed and the unique perpendicular speed, respectively. $v_r$ and $u_r$ are parameters that define the parallel and perpendicular velocity spread of the fast ions, respectively.

In figure 1, we compare the output of BO program with both linear analytical theory and the linear stages of first principles fully kinetic nonlinear simulations for MCI excited by super-Alfvénic fast ions (alpha-particle). The simulation parameters of the BO program are the same as those in [59], where bulk plasma parameters approximate the outer midplane edge conditions of the JET Preliminary Tritium Experiment (PTE) pulse 26148. The magnetic field is 2.1T, electrons and bulk deuterons are thermalized at 1keV. The bulk deuteron number density $n_D$ is $10^{19} \text{m}^{-3}$, and the number density ratio $\xi_F = n_F/n_D = 10^{-3}$, where $n_F$ represents fast ion number density. The rest of the simulation parameters are that $u_\perp = 1.294 \times 10^7 \text{ m/s}$, $v_d = 0 \text{ m/s}$, $v_r = u_r = 0 \text{ m/s}$, the Alfvén velocity $v_A = 1.02 \times 10^7 \text{ m/s}$, and the propagation angle $\theta = 88°$. From figure 1, it can be seen that the simulation results obtained using the BO program are basically consistent with those from the nonlinear simulations and linear analytical theory.

Figure 2 shows the comparison between the BO program and the variant of the linear MCI theory regarding the growth rate of instability excited by greatly sub-Alfvénic fast ions (deuteron). The simulation parameters used in the BO program are, following Ref. [54], magnetic field $B_0 = 5\text{T}$, bulk deuteron temperature $T_D = 3\text{keV}$, electron temperature $T_e = 1.5\text{keV}$, bulk deuteron number density $n_D = 1.5 \times 10^{19} \text{m}^{-3}$, the number density ratio $\xi_F = n_F/n_D = 10^{-2}$, $u_\perp = 2.4 \times 10^6 \text{ m/s}$, $v_d = 2.4 \times 10^6 \text{ m/s}$, $v_r = 2.4 \times 10^4 \text{ m/s}$, $u_r = 0 \text{ m/s}$, $v_A = 1.991 \times 10^7 \text{ m/s}$, and $\theta = 88.5°$. From figure 2, the BO simulation results are in good agreement with the linear theory.

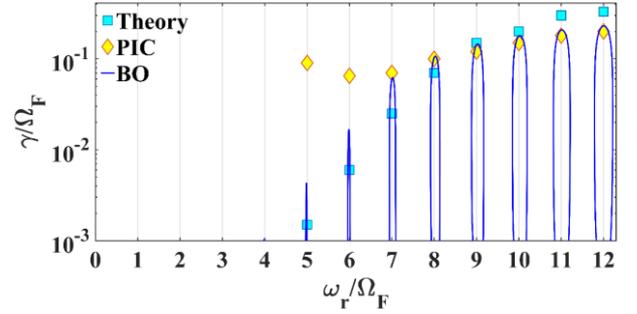

Figure 1 Growth rate of analytical linear theory from figure 1(b) of [59], nonlinear simulation from figure 1(b) of [59], and BO simulation for MCI excited by super-Alfvénic fast ions in cyclotron harmonics up to 12. $\Omega_F$ is the cyclotron frequency of fast ion.

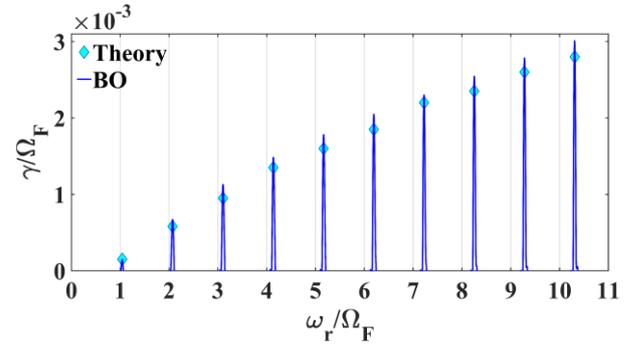

Figure 2 Analytical linear growth rate from figure 3 of [54] along with corresponding results from BO simulations for instability excited by greatly sub-Alfvénic fast ions.

Through the successful verification of MCI excited by super-Alfvénic and greatly sub-Alfvénic fast ions, the maturity and reliability of the BO program for MCI simulations have been demonstrated. Next, we will conduct detailed simulation on the velocity spread of the fast ions, the number density ratio, and the instability propagation angle for the cases of the MCI excited by super-Alfvénic, sub-Alfvénic, and greatly sub-Alfvénic fast ions, respectively. It is important to note the classification of fast ions into these categories [4]. On the one hand, the physical mechanism of ICE excited by greatly sub-Alfvénic fast ions differs from that of ICE excited by super-Alfvénic and sub-Alfvénic fast ions. On the other hand, the excitation conditions of MCI are distinctly different for the above three fast ions. In general, the super-Alfvénic fast ions can drive the MCI even if they are isotropic or have a relatively broad distribution of speeds[4]. The sub-Alfvénic fast ions that are isotropic or have undergone a certain degree of thermalization cannot drive the MCI[4]. For the greatly sub-Alfvénic fast ions with a very narrow spread of velocities in the parallel direction the instability can occur[4]. In addition, the above three ions show more differences in this work.

## 3. Super-Alfvénic fast ions

In the JET Preliminary Tritium Experiment, ICE excited by super-Alfvénic fast ions (alpha-particle) exhibits numerous features, and linear and nonlinear simulations have captured and explained most of the essential features [4,53,56,58-60]. However, the high cyclotron harmonic range has been seldom considered in the previous simulations. Based on this, the present work studies that the computational spectra region is up to 40 times of the ion cyclotron frequency while a more comprehensive MCI simulation results on the velocity spread of the fast ions, the number density ratio, and the instability propagation angle are presented. Additionally, we consider a more realistic experimental condition by including a certain percentage of tritium in the background plasma for deuterium-tritium plasma simulations. The simulated parameters are consistent with those used in section 2 regarding MCI excited by super-Alfvénic fast ions.

### 3.1 Velocity spread

The velocity spread of fast ions generated by D-T fusion is influenced by the rise time of neutron emission and the fast ion slowing-down time. The research indicates that if the rise time of neutron emission exceeds the fast ion slowing-down time, collisions will cause the energy distribution of fast ions to broaden before new fast ions are added[89-91]. Therefore, it is necessary to consider the impact of energy distribution on ICE. Studies on the effects of velocity spread on ICE have explored both the impact of considering only parallel velocity spread [50,57] and the combined effects of parallel and perpendicular velocity spread [2,55,70] , which corresponds to the case of isotropic temperature. These studies have all shown a suppressive effect on ICE. In our simulation, we consider isotropic temperature, i.e., $v_r = u_r$, with the simulation range from 0 to $0.4u_\perp$. Figures 3 and 4 show that the growth rates of cyclotron harmonics up to $l = 40$ are plotted as a function of $\omega_r$ for velocity spread ranging from 0 to $0.4u_\perp$. The results show that the growth rates of MCI gradually decrease with increasing $v_r$ and $u_r$ when the cyclotron harmonics are relatively low, i.e., less than 17, which is consistent with the previous simulation results (c.f., [55]). However, when considering higher harmonics, the simulations on MCI reveal more unique phenomena. Firstly, when $v_r = u_r < 0.1u_\perp$, the harmonics greater than 18 are divided into four intervals, with the centers of these intervals located at $l = 25, 32, 36,$ and 39, respectively. Secondly, the harmonics, specifically $l = 19, 20, 21,$ and 29, are obviously suppressed in $v_r = u_r = 0$. Thirdly, in terms of the high harmonics, there is a significant suppression as $v_r$ and $u_r$ increase when $v_r = u_r < 0.1u_\perp$. Overall, the growth rate of most harmonics decreases with increasing $v_r$ and $u_r$. However, for boundary harmonics such as $l = 22$, in the four intervals, the growth rate decreases sharply first, then increases slightly, and finally decreases gradually with increasing $v_r$ and $u_r$. Hence, the previous rule does not apply to such harmonics.

A particularly interesting phenomenon arises when $v_r$ and $u_r$ exceed $0.22u_\perp$: a continuous spectrum forms at the high harmonics, representing the first instance in which a simulation captures a continuous spectrum resembling experimental results. To understand the factors contributing to this phenomenon, we conduct a separate investigation into the influence of $v_r$ and $u_r$ on MCI. Figure 5 shows that the growth rates of cyclotron harmonics are plotted as a function of $\omega_r$ for (a)$v_r = 0.4u_\perp$, $u_r = 0$, and (b)$v_r = 0$, $u_r = 0.4u_\perp$. By comparing Figure 5 with Figure 3(a), we can clearly see that $v_r$ exerts a minor suppressive effect on the growth rate of MCI but plays a decisive role in determining the bandwidth of harmonics and the presence of the continuous spectrum. Conversely, $u_r$ significantly influences the growth rate of MCI. It should be noted that it is understandable that the previous simulations did not capture this key feature. In previous linear simulations, $v_r$ was consistently small, and only a single low harmonic was considered. In previous nonlinear simulations, where the excitation energy of MCI stems from perpendicular speed, $v_r$ remained small during the nonlinear evolution. Thus, in future nonlinear simulations containing a greater $v_r$, a continuum spectrum is foreseen.

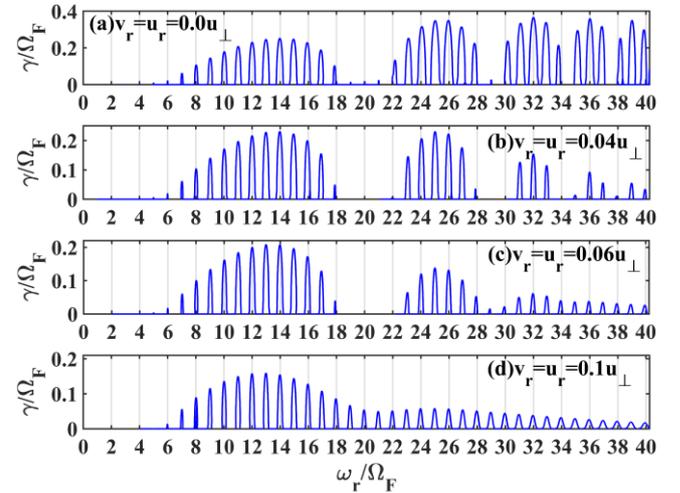

Figure 3 Growth rate of MCI excited by super-Alfvénic fast ions as a function of $\omega_r$ for (a)$v_r = u_r = 0$, (b)$v_r = u_r = 0.04u_\perp$, (c)$v_r = u_r = 0.06u_\perp$, and (d)$v_r = u_r = 0.1u_\perp$.

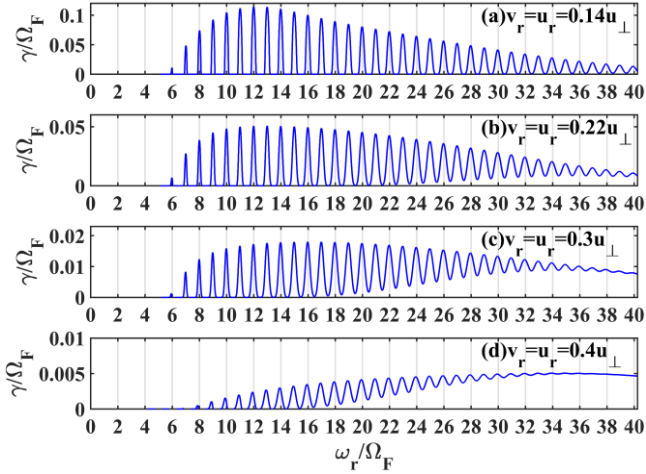

Figure 4 Growth rate of MCI excited by super-Alfvénic fast ions as a function of $\omega_r$ for (a) $v_r = u_r = 0.14u_\perp$, (b) $v_r = u_r = 0.22u_\perp$, (c) $v_r = u_r = 0.3u_\perp$, and (d) $v_r = u_r = 0.4u_\perp$.

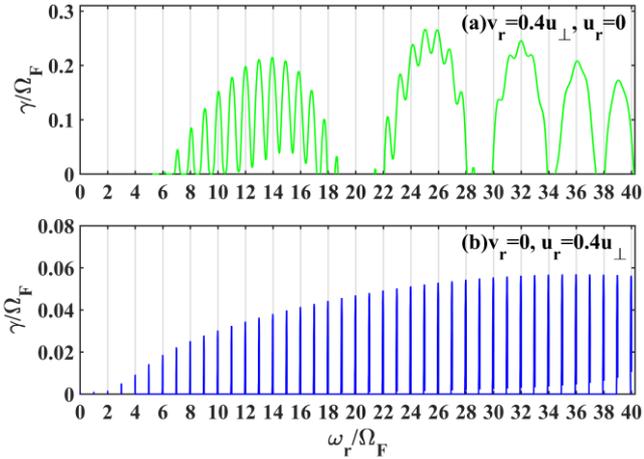

Figure 5 Growth rate of MCI excited by super-Alfvénic fast ions as a function of $\omega_r$ for (a) $v_r = 0.4u_\perp$, $u_r = 0$, and (b) $v_r = 0$, $u_r = 0.4u_\perp$.

### 3.2 Number density ratio

In JET [1], a linear relation was found between ICE intensity and neutron rate over six orders of magnitude. This observation not only suggests that ICE is excited by fast ions but also indicates a strong connection between fast ion number density and MCI. The simulations have successfully reproduced this experimental phenomenon and revealed a linear relationship between $\sqrt{\xi_F}$ and the MCI growth rate, with $\xi_F$ spanning 2 to 3 orders of magnitude in the linear phase [58,60]. In our simulations, our study explores a wider range of $\xi_F$ spanning 5 orders of magnitude: $\xi_F = 5 \times 10^{-7}$, $10^{-6}$, $10^{-5}$, $10^{-4}$, and $10^{-3}$, while keeping other simulation parameters constant. Figure 6 shows the ratio of the MCI growth rate $\gamma$ to $\sqrt{\xi_F}$ versus $\xi_F$, where a straight horizontal trend implies a relationship $\gamma \sim \sqrt{\xi_F}$. It should be noted that the properties of the four high harmonic intervals regarding the relation between $\gamma$ and $\sqrt{\xi_F}$ is similar. Therefore, figure 6 only displays the relationship between $\gamma$ and $\sqrt{\xi_F}$ for each harmonic within one harmonic range ($22 \leq l \leq 28$). Comparing figure 6 and figure 7(a), we observe that these high harmonics conforming to the linear relation appear only in the centers of the four intervals. In addition, when $l < 22$, the growth rate of all harmonics except $l = 17$, $l = 18$, and $l < 6$ varies linearly with $\sqrt{\xi_F}$, which is consistent with the previous simulation results [60]. Figure 7 shows the growth rates of cyclotron harmonics up to $l = 40$ are plotted as a function of $\omega_r$ for $\xi_F$ ranging from $10^{-6}$ to $10^{-3}$. It is evident that as $\xi_F$ decreases, the growth rate of each harmonic gradually decreases. Similar to $v_r$, $\xi_F$ also exerts a significant influence on the bandwidth of harmonics.

Lastly, considering the self-limitation of MCI, it is reasonable to infer that the linear relationship between ICE intensity and neutron rate during the nonlinear stage is not unrelated to the linear stage. Therefore, it is natural to speculate that the velocity spread plays a crucial role in the relationship between $\gamma$ and $\xi_F$. Figure 8 shows $\gamma/\Omega_F$ versus $\xi_F$ at different velocity spreads. The figure illustrates that the relationship $\gamma/\Omega_F \sim \sqrt{\xi_F}$ (figure 8 (a)) transitions to $\gamma/\Omega_F \sim \xi_F$ (figure 8 (d)) with increasing the velocity spread. Hence, this work confirms that the linear relationship between ICE intensity and neutron rate may be determined by a linear mechanism.

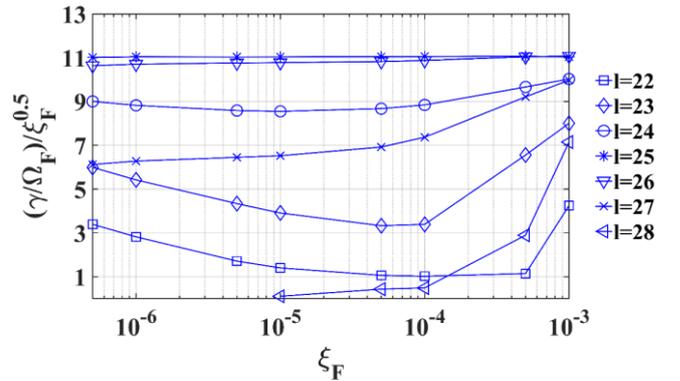

Figure 6 Ratio of growth rate to $\sqrt{\xi_F}$ as a function of $\xi_F$ for the harmonic number $l = 22$, $23$, $24$, $25$, $26$, $27$, and $28$.

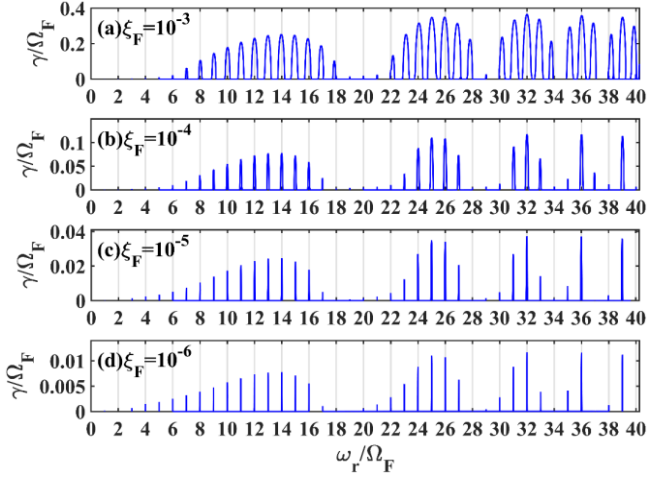

Figure 7 Growth rate of MCI excited by super-Alfvénic fast ions as a function of $\omega_r$ for (a)$\xi_F = 10^{-3}$, (b)$\xi_F = 10^{-4}$, (c)$\xi_F = 10^{-5}$, and (d)$\xi_F = 10^{-6}$.

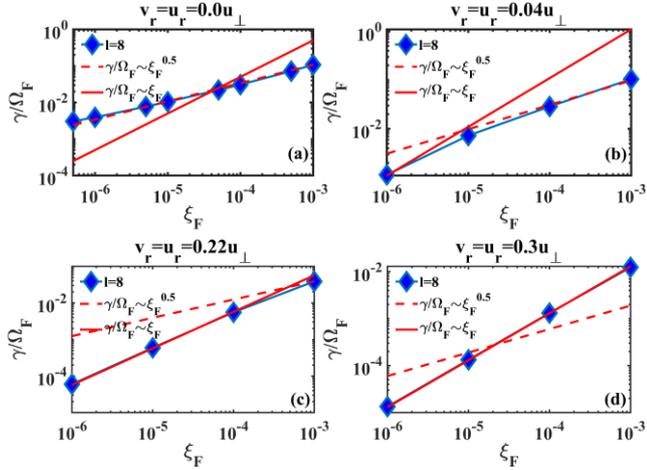

Figure 8 For super-Alfvénic fast ions instability growth rate $\gamma/\Omega_F$ as a function of the number density ratio $\xi_F$, calculated for the harmonic number $l = 8$ at different velocity spreads. The red dashed and solid red lines correspond to $\gamma/\Omega_F \sim \sqrt{\xi_F}$ and $\gamma/\Omega_F \sim \xi_F$, respectively. The solid blue line shows the result of a numerical calculation.

*3.3 Propagation angle*

In the development of linear MCI theory, the fast Alfvén and ion Bernstein branches is extended from perpendicular to oblique propagation. While Landau damping and ion cyclotron damping have been introduced into MCI, it has been observed through both linear and nonlinear simulations [53,56,57,62] that all ion cyclotron harmonics are excited simultaneously, exhibiting a strong growth rate for nearly perpendicular wave propagation. However, previous linear simulations neglected the wave electric field parallel to the ambient magnetic field, limiting the strict self-consistency of the results to large instability propagation angles. In nonlinear simulations, the instability propagation angle is also limited to a small range close to perpendicular direction. Therefore, we use the BO program, which includes the wave electric field parallel to the ambient magnetic field, to simulate the effect of a large angle deviating from the perpendicular direction on MCI excitation. We consider a large angle of deviation ($15° \leq \theta \leq 90°$) and account for the parallel drift speed $v_d$, which cannot be neglected in such cases. Based on [1,53], we set $v_r = u_r = 0.04u_\perp$ and $v_d = 0.25u_\perp$. Figure 9 shows that the growth rates of cyclotron harmonics up to $l = 40$ are plotted as a function of $\omega_r$ for $\theta$ ranging from $50°$ to $88°$. We do not present simulations with propagation angles greater than $90°$ since the forward and reverse propagating waves of angles greater than $90°$ are equivalent to the reverse and forward propagating waves of angles less than $90°$, respectively. Overall, as the propagation angle decreases, almost all harmonics are basically suppressed at propagation angles $\theta < 15°$. At about $85°$, the harmonics $|l| \geq 15$ are almost suppressed, indicating that the high harmonics are more sensitive to the propagation angle than low harmonics.

The rapid suppression of MCI with increasing the angle of deviation from the perpendicular direction is related to the dispersion relation of the fast Alfvén wave. Figure 10 shows the dispersion relation of the fast Alfvén wave at different angles, where the straight lines and curves correspond to ion Bernstein and fast Alfvén waves, respectively. The dispersion relation changes when the propagation angle changes from $88°$ to $80°$, which causes $u_\perp$ to become smaller than the perpendicular phase velocity of the fast Alfvén wave $v_{A\perp}$ at harmonics $|l| > 10$, invalidating the super-Alfvénic condition and making the sub-Alfvénic condition valid. This eventually leads to the suppression of the harmonics. It needs to be noted that previous references focused on the case of low harmonic numbers and large propagation angles, where the dispersion relationship of linear Alfvén waves $\omega = k_\perp v_A$ can describe the dispersion relationship of Alfvén waves well. Therefore, comparing $u_\perp$ with the Alfvén speed $v_A$ is meaningful. However, when the harmonic number is large, the dispersion relationship of Alfvén waves changes, and it makes sense to compare the $u_\perp$ with the perpendicular phase velocity of fast Alfvén wave $v_{A\perp}$, as defined in this work.

One important observation from figure 9 is that each harmonic consists of both forward and reverse propagating waves, with the forward wave having a higher frequency than the reverse wave. The line splitting, which is evident in figure 9 due to $\theta$ and $v_d$, provides a simple explanation [4,53] for the spectral peak splitting observed in JET experiments. Furthermore, recent computational results reported in Ref. [85] indicate that the origin of spectral peak splitting is Doppler-shifted resonances and the intricate landscape of the MCI growth rate on the dispersion surface in $(k_\perp, k_\parallel)$ space.

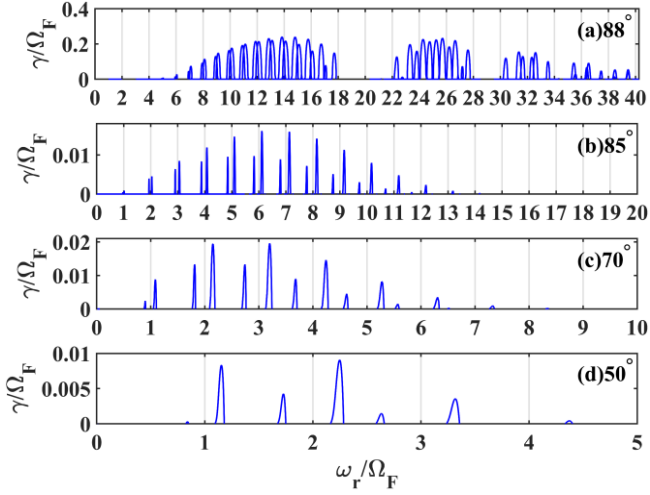

Figure 9 Growth rate of MCI excited by super-Alfvénic fast ions as a function of $\omega_r$ for (a)$\theta = 88°$, (b)$\theta = 85°$, (c)$\theta = 70°$, and (d)$\theta = 50°$.

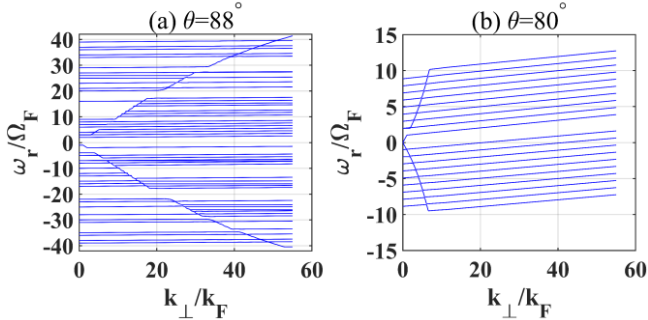

Figure 10 Dispersion relation of fast Alfvén wave for (a)$\theta = 88°$, (b)$\theta = 80°$. The straight lines and curves correspond to ion Bernstein and fast Alfvén wave, respectively. Here, $k_F = \Omega_F/v_A$.

### 3.4 Containing tritium

In previous simulations of the JET Preliminary Tritium Experiment [50,53,55-60], the background plasma was typically assumed to be deuterium plasma, which is a reasonable approximation for low tritium number density. However, for the higher tritium-to-deuterium number density ratios expected in the future ITER, it is necessary to contain a corresponding percentage of tritium in the background plasma. Specifically, we set the maximum ratio of tritium to total ion number density to 0.3, while keeping other parameters unchanged. Figure 11 shows that the growth rates of cyclotron harmonics up to $l = 40$ are plotted as a function of $\omega_r$ for tritium number density ratio ranging from 0 to 0.3. It is clear from the figure that, as the tritium number density ratio increases, the four high harmonic intervals move toward the low harmonics. The harmonics that are significantly suppressed not only exhibit the same trend, but also experience a slight increase in their number.

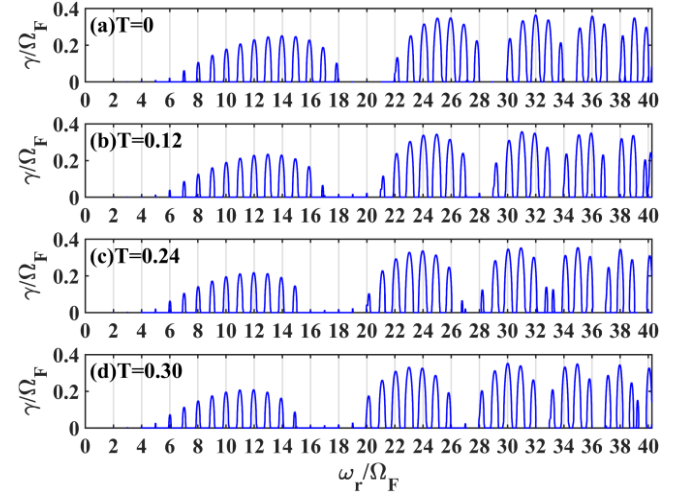

Figure 11 Growth rate of MCI excited by super-Alfvénic fast ions as a function of $\omega_r$ for (a)T = 0, (b)T = 0.12, (c)T = 0.24, and (d)T = 0.30.

## 4. Sub-Alfvénic fast ions

ICEs excited by sub-Alfvénic fast ions are commonly observed and have been widely studied through simulations. Here, as with the simulation of the MCI excited by super-Alfvénic fast ions, we conduct a comprehensive simulation of the key parameters for MCI excited the sub-Alfvénic fast ions, with consideration for cyclotron harmonics up to 40. The simulation parameters obtained from LHD are, following Ref. [62], magnetic field $B_0 = 0.46\text{T}$, bulk proton temperature $T_H = 150\text{eV}$, electron temperature $T_e = 150\text{eV}$, bulk proton number density $n_H = 10^{19}\text{m}^{-3}$, the ratio of the fast ion (proton) number density to the background proton number density $\xi_F = n_F/n_H = 5 \times 10^{-4}$, $u_\perp = 2.77 \times 10^6 \text{ m/s}$, $v_d = 0 \text{ m/s}$, $v_r = u_r = 0 \text{ m/s}$, $v_A = 3.17 \times 10^6 \text{ m/s}$, and $\theta = 89°$.

### 4.1 Velocity spread

As mentioned in section 2, ICEs excited by sub-Alfvénic fast ions depend more on the velocity spread of the fast ions than those excited by super-Alfvénic fast ions. However, this conclusion is obtained based on low harmonics, and new results have emerged with the inclusion of high harmonics in the present study. Figure 12, where the blue lines represent $l \leq 28$ and the red lines represent $l \geq 29$, show that the growth rates of cyclotron harmonics up to $l = 40$ are plotted as a function of $\omega_r$ for velocity spread ranging from 0 to $0.35u_\perp$. The figure shows that the growth rate of harmonics less than 29 is one to two orders of magnitude smaller than those of the harmonics greater than 29, indicating that high harmonics are more likely to be excited. As the velocity spread increases, the harmonics less than 29 are rapidly suppressed, consistent with the previous simulation results that only consider the velocity spread in the perpendicular direction [64] or the combined effects of velocity spread in parallel and perpendicular [55], i.e., the isotropic temperature case, which all show a suppressive effect on ICE. The growth rate of harmonics greater than 28 gradually decreases, while continuous spectrum features appear, similar to the MCI excited by super-Alfvénic fast ions. Further analysis reveals that the characteristics observed in high harmonics are related to the dispersion relation of the fast Alfvén wave. Figure 13 shows that $v_{A\perp}$ decreases gradually with the increase of harmonics, eventually resulting in $u_\perp > v_{A\perp}$. Consequently, the sub-Alfvénic condition becomes invalid at harmonics above 28 (the black circle mark in the figure), while the super-Alfvénic condition becomes valid, resulting in a drastic increase in the growth rate at harmonics greater than 28. Finally, it should be noted that from the figure 12(a) the simulation results of the BO differ significantly from those in [62] at $l = 11$. This is because the wave-wave coupling, which leads to an increase in the growth rate [59], occurs in the late linear phase of the simulations in [62] while the results of the BO here regarding the MCI excited by sub-Alfvénic fast ions are in the early linear phase.

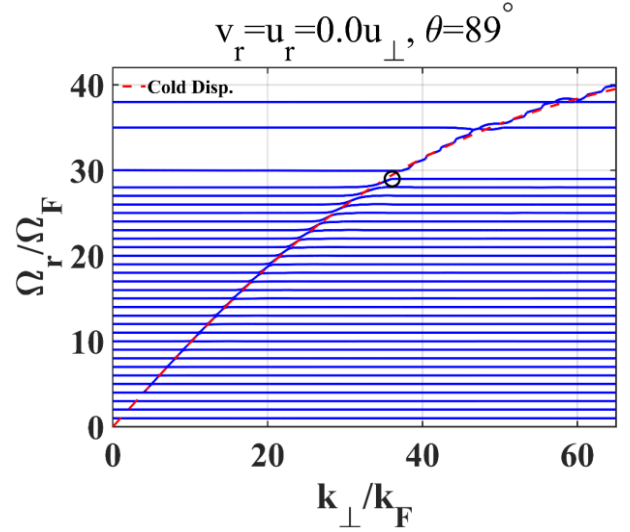

Figure 13 Dispersion relation of fast Alfvén wave. The blue straight lines and blue curves correspond to ion Bernstein and fast Alfvén wave, respectively. The cold plasma dispersion relation for the fast Alfvén wave, is calculated by the fluid module of the BO program, is shown by the red dashed line.

### 4.2 Number density ratio

Similar to the MCI excited by super-Alfvénic fast ions, detailed simulations about the number density ratio $\xi_F$ is presented for the MCI excited by sub-Alfvénic fast ions. The simulation sets $\xi_F = 10^{-6}$, $10^{-5}$, $10^{-4}$, and $10^{-3}$, while keeping other parameters unchanged. Figure 14 illustrates that as the number density ratio $\xi_F$ decreases, the growth rate of each harmonic also decreases. Moreover, for harmonics above 29, they are divided into three intervals centered around $l = 33$, 37, and 39, respectively. We find that these high harmonics, which exhibit a linear relationship between $\sqrt{\xi_F}$ and $\gamma$, appear only at the centers of the three intervals, as shown in figure 15(b), where the interval encompassing $l = 36$, 37 and 38 is depicted. Therefore, the high harmonics with the typical characteristics resembling MCI excited by super-Alfvénic fast ions further supports the conclusions presented in section 4.1. Lastly, for harmonics less than 29, such as $l = 15$ in figure 15(a), the growth rate $\gamma$ is close to linear relation with $\sqrt{\xi_F}$, which is consistent with the previous results [62].

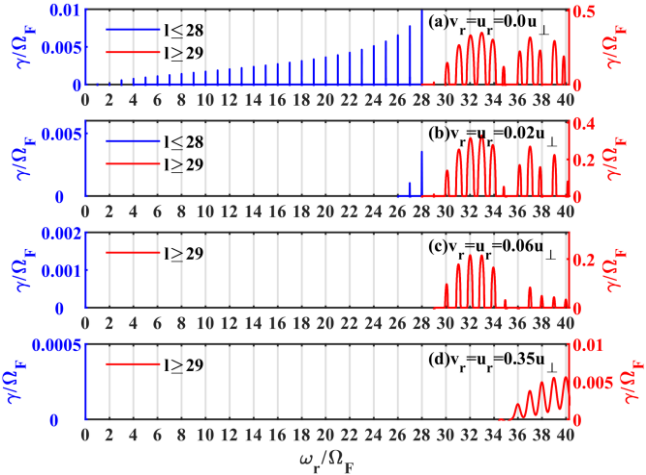

Figure 12 Growth rate of MCI excited by sub-Alfvénic fast ions as a function of $\omega_r$ for (a)$v_r = u_r = 0.0u_\perp$, (b)$v_r = u_r = 0.02u_\perp$, (c)$v_r = u_r = 0.06u_\perp$, (d)$v_r = u_r = 0.35u_\perp$. The blue lines represent $l \leq 28$ , and its ordinate is on the left side of the figure. The red lines represent $l \geq 29$, and its ordinate is on the right side of the figure.

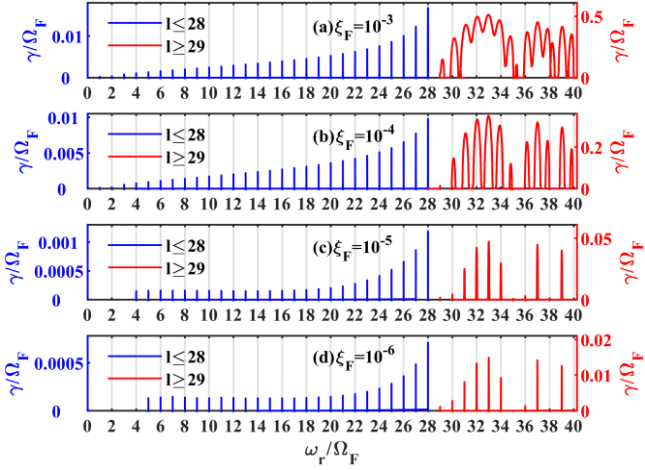

Figure 14 Growth rate of MCI excited by sub-Alfvénic fast ions as a function of $\omega_r$ for (a)$\xi_F = 10^{-3}$, (b)$\xi_F = 10^{-4}$, (c)$\xi_F = 10^{-5}$, and (d)$\xi_F = 10^{-6}$. The blue lines represent $l \leq 28$, and its ordinate is on the left side of the figure. The red lines represent $l \geq 29$, and its ordinate is on the right side of the figure.

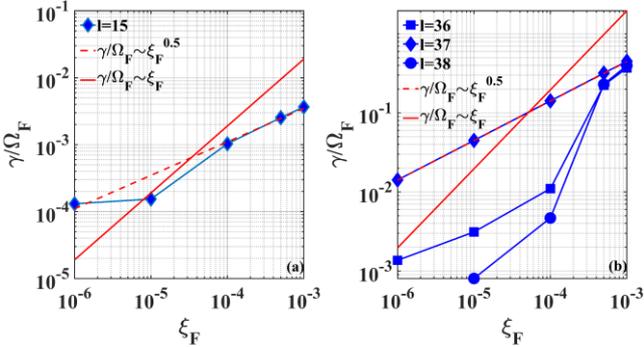

Figure 15 For sub-Alfvénic fast ions instability growth rate $\gamma/\Omega_F$ as a function of the number density ratio $\xi_F$, calculated for (a) $l = 15$ and (b) $l = 36$, 37 and 38. The red dashed and solid red lines correspond to $\gamma/\Omega_F \sim \sqrt{\xi_F}$ and $\gamma/\Omega_F \sim \xi_F$, respectively. The solid blue line shows the result of a numerical calculation.

### 4.3 Propagation angle

In our research on the relationship between the MCI excited by super-Alfvénic fast ions and the propagation angle, simulation results indicate that the larger the angle of deviation from the perpendicular direction, the stronger the suppression of high harmonics. We have explored the relationship between MCI excited by sub-Alfvénic fast ions and the propagation angle, and the conclusions at the high harmonics are consistent with those of the MCI excited super-Alfvénic fast ions. From figure 16, which shows the relation between MCI growth rate and propagation angle, we can see that the high harmonics are first strongly suppressed as the angle of deviation from the perpendicular direction increases. Near $\theta = 85°$, the growth rate of high harmonics is in the same order of magnitude with that of low harmonics, indicating that the super-Alfvénic condition becomes invalid at harmonics above 28, while the sub-Alfvénic condition becomes valid. As the propagation angle decreases further, the growth rate of high harmonics first gradually decreases. However, the growth rate of low harmonics does not decrease monotonically with decreasing propagation angle, but increase first and then decrease, which is consistent with the previous results [66]. When the propagation angle is less than $10°$, all harmonics are basically suppressed.

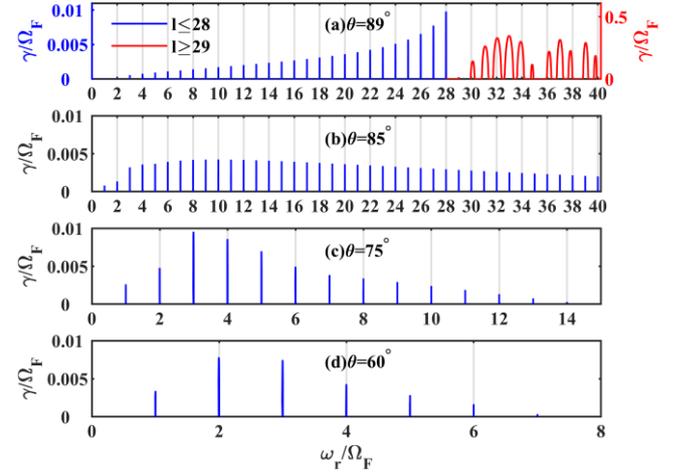

Figure 16 Growth rate of MCI excited by sub-Alfvénic fast ions as a function of $\omega_r$ for (a)$\theta = 89°$, where the blue lines, whose ordinate is on the left side of the figure, represent $l \leq 28$, and the red lines, whose ordinate is on the right side of the figure, represent $l \geq 29$, (b)$\theta = 85°$, (b)$\theta = 75°$, and (d)$\theta = 60°$.

## 5. Greatly sub-Alfvénic fast ions

Different from the electromagnetic instability excited by super-Alfvénic and sub-Alfvénic fast ions, the instability excited by greatly sub-Alfvénic fast ions, which is a variant of MCI [54], is mainly electrostatic. There have been relatively few simulation studies on the ICE excited by greatly sub-Alfvénic fast ions. Here we conducted a comprehensive simulation study on MCI, taking into account the velocity spread of the fast ions (deuteron), the number density ratio, and the instability propagation angle. The simulation parameters are, following Ref. [54,92], bulk deuteron temperature $T_D = 4\text{keV}$, and electron temperature $T_e = 3\text{keV}$. Other parameters remain the same as those used for greatly sub-Alfvénic fast ions discussed in section 2. In our subsequent simulations, we focus solely on the forward propagating waves, as the forward and reverse propagating waves exhibit similar properties, and this choice enhances the clarity of the presented results. In addition, we compare the electromagnetic and electrostatic results carried out by using the program BO. Figure 17 illustrates that the results are largely consistent, supporting the analytical conclusion [54].

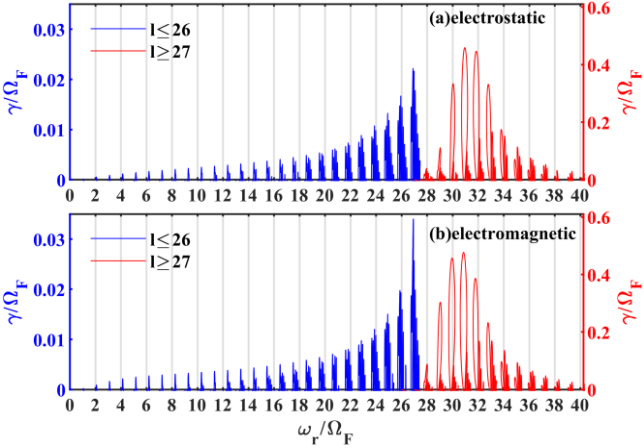

Figure 17 (a)electrostatic results along with corresponding (b)electromagnetic results for instability excited by greatly sub-Alfvénic fast ions. The blue lines represent $l \leq 26$, and its ordinate is on the left side of the figure. The red lines represent $l \geq 27$, and its ordinate is on the right side of the figure.

### 5.1 Velocity spread

As mentioned earlier, the excitation of ICE by greatly sub-Alfvénic fast ions requires a very narrow spread of velocities in the parallel direction [4,54], on the order of $10^{-2}u_\perp$. The simulation results of the BO are consistent with the previous simulation results at low harmonics. However, when considering high harmonics, the excitation condition for MCI by greatly sub-Alfvénic fast ions becomes relatively relaxed. Figure 18 shows that the growth rates of cyclotron harmonics up to $l = 40$ are plotted as a function of $\omega_r$ for velocity spread ranging from 0 to $0.1u_\perp$. The figure shows that MCI excited by greatly sub-Alfvénic fast ions exhibits a similar characteristic to that excited by sub-Alfvénic fast ions at high harmonics, where the growth rate of high harmonics is one to two orders of magnitude larger than that of low harmonics. Different from the MCI excited by super-Alfvénic and sub-Alfvénic fast ions, the MCI excited by the greatly sub-Alfvénic fast ions shows more line splitting (the line splitting of high harmonics in figure 18(a) is similar to that of low harmonics, but the details are not shown due to the relatively small values of the finite structure compared to its maximum value), which can be well understood through equation (11) in Ref. [54]. Another important characteristic is that with increasing velocity spread, low harmonics are rapidly suppressed, while a few high harmonics still persist when $v_r = u_r = 0.1u_\perp$. This expands the parameter range for studying ICE excited by greatly sub-Alfvénic fast ions and is significant for experimental investigations of ICE.

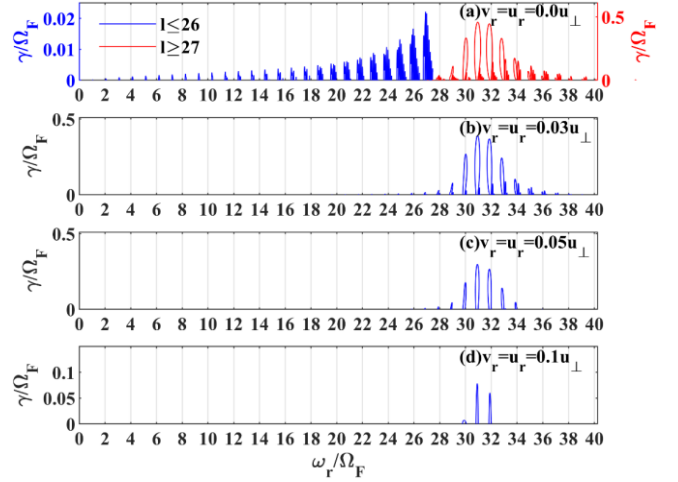

Figure 18 Growth rate of MCI excited by greatly sub-Alfvénic fast ions as a function of $\omega_r$ for (a)$v_r = u_r = 0$, where the blue lines, whose ordinate is on the left side of the figure, represent $l \leq 26$, and the red lines, whose ordinate is on the right side of the figure, represent $l \geq 27$, (b)$v_r = u_r = 0.03u_\perp$, (c)$v_r = u_r = 0.05u_\perp$, and (d)$v_r = u_r = 0.1u_\perp$.

### 5.2 Number density ratio

In the TFTR experiment, the number density ratio $\xi_F$ is on the order of $10^{-2}$, and relevant simulations have shown that exciting MCI for low harmonics is difficult at lower number density ratios [54]. Here we further study the influence of number density ratio on each harmonic, especially the high harmonics. Figure 19 plots the growth rates of cyclotron harmonics up to $l = 40$ as a function of $\omega_r$ for the number density ratio $\xi_F$ ranging from $10^{-4}$ to $10^{-1}$. From the figure, we can see that the low harmonics are suppressed at a lower number density ratio. However, even when $\xi_F$ is reduced to $10^{-4}$, a high harmonic persists. In addition, at $\xi_F = 10^{-1}$, a continuous spectrum is formed at the high harmonics. Overall, as $\xi_F$ decreases, the growth rates of both low and high harmonics decrease rapidly, accompanied by narrower bandwidths. This behavior is similar to that of super-Alfvénic and sub-Alfvénic fast ions.

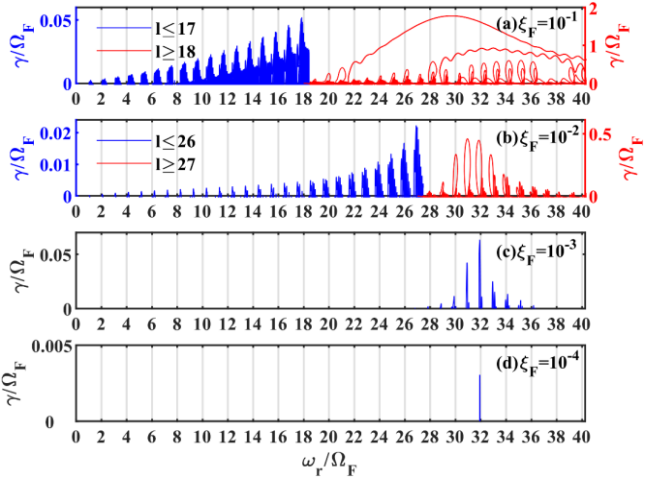

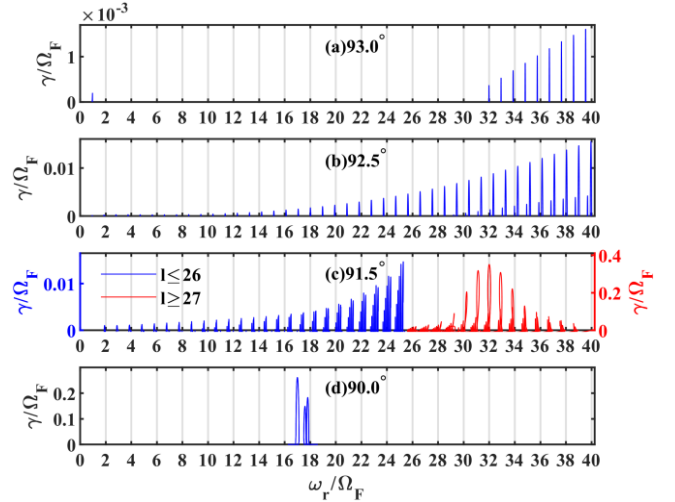

Figure 19 Growth rate of MCI excited by greatly sub-Alfvénic fast ions as a function of $\omega_r$ for (a)$\xi_F = 10^{-1}$, where the blue lines, whose ordinate is on the left side of the figure, represent $l \leq 17$, and the red lines, whose ordinate is on the right side of the figure, represent $l \geq 18$, (b)$\xi_F = 10^{-2}$, where the blue lines, whose ordinate is on the left side of the figure, represent $l \leq 26$, and the red lines, whose ordinate is on the right side of the figure, represent $l \geq 27$, (c)$\xi_F = 10^{-3}$, and (d)$\xi_F = 10^{-4}$.

*5.3 Propagation angle*

Here we study in detail the relationship between the instability excited by greatly sub-Alfvénic fast ions and the propagation angle. Figures 20 and 21 show that the growth rates of cyclotron harmonics up to $l = 40$ are plotted as a function of $\omega_r$ for propagation angle ranging from $87°$ to $93°$. Notably, the instability excited by greatly sub-Alfvénic fast ions is more sensitive to the propagation angle than the instability excited by super-Alfvénic and sub-Alfvénic fast ions, and is basically suppressed when the propagation angle deviates from the perpendicular direction by about $4°$. Overall, the growth rate of each harmonic is strong for nearly perpendicular propagation. As the angle of deviation from the perpendicular direction increases, the most unstable harmonic remains in the high harmonic range, while the harmonics at the middle harmonic range are the first to be suppressed. This is different from the instability excited by super-Alfvénic and sub-Alfvénic fast ions, where the high harmonics are typically suppressed first.

Figure 20 Growth rate of MCI excited by greatly sub-Alfvénic fast ions as a function of $\omega_r$ for (a)$\theta = 93°$, (b)$\theta = 92.5°$, (c)$\theta = 91.5°$, where the blue lines, whose ordinate is on the left side of the figure, represent $l \leq 26$, and the red lines, whose ordinate is on the right side of the figure, represent $l \geq 27$, and (d)$\theta = 90°$.

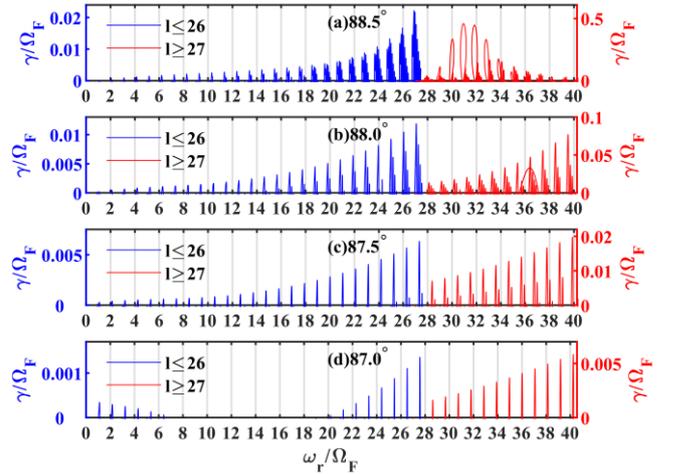

Figure 21 Growth rate of MCI excited by greatly sub-Alfvénic fast ions as a function of $\omega_r$ for (a)$\theta = 88.5°$, (b)$\theta = 88°$, (c)$\theta = 87.5°$, and (d)$\theta = 87°$. The blue lines represent $l \leq 26$, and its ordinate is on the left side of the figure. The red lines represent $l \geq 27$, and its ordinate is on the right side of the figure.

## 6. Summary and Conclusions

In this work, we systematically investigate the effects of key parameters on linear MCI in three cases: super-Alfvénic, sub-Alfvénic, and greatly sub-Alfvénic fast ions with drift ring beam distribution. Our simulation results are consistent with the previous ones, which are summarized in table 1, in the low harmonic range. However, when expanding the computational spectra region up to 40 times of the ion cyclotron frequency, many new features appear. For MCI excited by super-Alfvénic fast ions, harmonics greater than 18 are divided into four intervals. For MCI excited by sub-Alfvénic and greatly sub-Alfvénic fast ions, the growth rate of high harmonics is one to two orders of magnitude larger than that of low harmonics. Additionally, our simulations on velocity spread, number density ratio, and instability propagation angle yield interesting results, summarized in table 2.

The first one is about the simulations of velocity spread. For MCI excited by super-Alfvénic fast ions, $v_r$ has a small suppressive effect on the growth rate but has a decisive effect on the bandwidth of harmonics and the continuous spectrum, while $u_r$ has a decisive effect on the growth rate. For MCI excited by sub-Alfvénic fast ions, the high harmonics form a continuous spectrum at a greater velocity spread. This indicates that the sub-Alfvénic condition becomes invalid above harmonics 28, while the super-Alfvénic condition becomes valid. For MCI excited by greatly sub-Alfvénic fast ions, the parameter range of the velocity spread, where fast ions can excite MCI, is expanded.

The second one is about the simulations of number density ratio. For MCI excited by the super-Alfvénic fast ions, an important conclusion is that the relationship $\gamma \sim \sqrt{\xi_F}$ transitions to $\gamma \sim \xi_F$ with increasing the velocity spread. This promotes the understanding of the linear relation of the fast ion number density with ICE intensity in the JET. For MCI excited by sub-Alfvénic fast ions, high harmonics conforming to the linear relation between $\sqrt{\xi_F}$ and $\gamma$ appear only in the centers of the three intervals, which shows the typical characteristics of MCI excited by super-Alfvénic fast ions. For MCI excited by greatly sub-Alfvénic fast ions, the parameter range of the number density ratio allowing fast ions to excite MCI is expanded.

The last one is about the simulations of propagation angle. For MCI excited by the super-Alfvénic and sub-Alfvénic fast ions, high harmonics are highly sensitive to the propagation angle compared with low harmonics. This is because the change in the dispersion relation of the fast Alfvén wave with propagation angle results in the transition of super-Alfvénic fast ions to sub-Alfvénic fast ions. In addition, low harmonics still persist at large angles of deviation from the perpendicular direction. For MCI excited by greatly sub-Alfvénic fast ions, the most unstable harmonic is still at the high harmonic range when the angle of deviation increases. The instability excited by the greatly sub-Alfvénic fast ions is highly sensitive to the propagation angle and is basically suppressed when the propagation angle deviates from the perpendicular direction by about 4°.

Lastly, we consider a more realistic experimental condition for MCI excited by super-Alfvénic fast ions, that is, the background plasma contains a certain percentage of tritium. The simulation results show that as the tritium number density ratio increases, the four high harmonic intervals move toward the low harmonics, and the harmonics that are significantly suppressed not only exhibit the same moving trend but also have a slight increase in their numbers.

In our current work, we have simulated the key parameters in detail on different devices such as JET (super-Alfvénic fast ions), LHD (sub-Alfvénic fast ions), TFTR (greatly sub-Alfvénic fast ions), and roughly summarized the rules. However, for different parameters of some different devices such as LHD (super-Alfvénic fast ions) [62], JET (sub-Alfvénic fast ions) [38], and TFTR (sub-Alfvénic fast ions) [55], we have also made corresponding simulations. The specific results may change for super-Alfvénic, sub-Alfvénic, and greatly sub-Alfvénic fast ions, but the rules are similar to those of the present work. For more wide parameters and more detailed results, the MCI simulation carried out by using the BO program is still anticipated. Finally, nonlinear simulations about the ICE continuous spectrum as well as wave-wave coupling in the linear phase would be interesting future works.

Table 1. Summary of the previous linear simulation results on MCI excited by super-Alfvénic fast ions, sub-Alfvénic fast ions, and greatly sub-Alfvénic fast ions, respectively, in low harmonic range.

| Parameter | Super-Alfvénic fast ions | Sub-Alfvénic fast ions | Greatly sub-Alfvénic fast ions |
|---|---|---|---|
| Velocity spread | MCI can occur if fast ions are isotropic or have a relatively broad distribution of speeds. | MCI cannot occur if fast ions are isotropic or a certain degree of thermalization. | MCI can occur under a very narrow spread of velocity in the parallel direction. |
| Number density ratio | $\gamma \sim \sqrt{\xi_F}$ | $\gamma \sim \sqrt{\xi_F}$ | MCI hard to excite at lower density ratios |
| Propagation angle | strong growth rate for nearly perpendicular wave propagation | nonmonotonically decreasing growth rate with decreasing propagation angle | |

Table 2. Summary of the new simulation results on MCI excited by super-Alfvénic fast ions, sub-Alfvénic fast ions, and greatly sub-Alfvénic fast ions, respectively.

| Parameter | Super-Alfvénic fast ions | Sub-Alfvénic fast ions | Greatly sub-Alfvénic fast ions |
|---|---|---|---|
| High harmonic range | four intervals | drastic increase in growth rate | drastic increase in growth rate |
| Velocity spread | $v_r$: determining continuous spectrum<br>$u_r$: determining growth rate | high harmonics: continuous spectrum similar to that of MCI excited super-Alfvénic fast ions | expanding parameter range |
| Number density ratio | $\gamma \sim \sqrt{\xi_F}$ transits to $\gamma \sim \xi_F$ with the increase of the velocity spread. | high harmonics: features similar to that of MCI excited super-Alfvénic fast ions | expanding parameter range |
| Propagation angle | high harmonics: sensitive to the propagation angle<br>low harmonics: existing in a small propagation angle | high harmonics: sensitive to the propagation angle<br>low harmonics: existing in a small propagation angle | All harmonics are suppressed at about $86°$. |

**Acknowledgments**

This work is supported by the National Natural Science Foundation of China under Grant No. 12275040, the National Key R&D Program of China under Grant Nos. 2019YFE03030004, Interdisciplinary and Collaborative Teams of CAS, and the High-End Talents Program of Hebei Province, Innovative Approaches towards Development of Carbon-Free Clean Fusion Energy (No. 2021HBQZYCSB006).

**Appendix A. BO model**

BO [86-88] (open source at https://github.com/hsxie/bo), which means 'wave' in Chinese, is a powerful solver that consists of two components: BO-F (PDRF), a multi-fluid solver, and BO-K (PDRK), a kinetic solver. BO-K effectively solves the uniform plasma dispersion relation with an extended Maxwellian based equilibrium distribution function, determining the solutions to $D(\omega, k) = 0$, given a specific wave vector $k$, by solving the series $\omega$.

One notable advantage of BO is its ability to solve the difficulty of root finding without requiring an initial guess, providing all important solutions simultaneously, with using of a novel matrix method. Additionally, BO supports various features, including anisotropic temperature, loss cone, drift in arbitrary direction, ring beam, collision, unmagnetized/magnetized species, electrostatic/electromagnetic/Darwin, and $k_\parallel \leq 0$. The BO code has been widely used in space and astrophysics plasmas (c.f., [93-95]) to study the abundant plasma waves and instabilities phenomena, where the uniform plasma assumption can be valid easily. A major purpose of the present work is also to show that it can be useful for study the waves and instabilities in fusion and laboratory plasmas. In addition, a ray tracing solver named BORAY[96], based on BO, for non-uniform plasmas has also been developed.